\documentclass[conference]{IEEEtran}
\IEEEoverridecommandlockouts
\usepackage[numbers]{natbib}
\setcitestyle{open={[},close={]}}

\usepackage{amsmath,amssymb,amsfonts}

\usepackage{graphicx}
\usepackage{textcomp}
\usepackage{xcolor}
\usepackage{float}
\usepackage{array}
\usepackage[font=footnotesize,labelfont=bf]{caption}
\usepackage{optidef}
\usepackage{subcaption}
\usepackage[utf8]{inputenc}
\def\BibTeX{{\rm B\kern-.05em{\sc i\kern-.025em b}\kern-.08em
    T\kern-.1667em\lower.7ex\hbox{E}\kern-.125emX}}

\usepackage{algorithm}
\usepackage{algpseudocode}
\usepackage{color, xcolor, colortbl}

\begin{document}

\title{Prediction and Detection of FDIA and DDoS Attacks in 5G Enabled IoT\\}
\author{\IEEEauthorblockN{
Hajar Moudoud \IEEEauthorrefmark{1}\IEEEauthorrefmark{3},
Lyes Khoukhi \IEEEauthorrefmark{2},
Soumaya Cherkaoui  \IEEEauthorrefmark{1},
}

\IEEEauthorblockA{
\IEEEauthorrefmark{1} Department of Electrical and Computer Engineering, Université de Sherbrooke, Canada\\ 
\IEEEauthorrefmark{2} GREYC CNRS, ENSICAEN, Normandie University, France\\
\IEEEauthorrefmark{3} University of Technology of Troyes, France \\ 
\{hajar.moudoud, soumaya.cherkaoui\}@usherbrooke.ca, lyes.khoukhi@ensicaen.fr}
}

\maketitle
\begin{abstract}
Security in the fifth generation (5G) networks has become one of the prime concerns in the telecommunication industry. 5G security challenges come from the fact that 5G networks involve different stakeholders using different security requirements and measures. Deficiencies in security management between these stakeholders can lead to security attacks. Therefore, security solutions should be conceived for the safe deployment of different 5G verticals (e.g., industry 4.0, Internet of Things (IoT), etc.). The interdependencies among 5G and fully connected systems, such as IoT, entail some standard security requirements, namely integrity, availability, and confidentiality. In this article, we propose a hierarchical architecture for securing 5G enabled IoT networks, and a security model for the prediction and detection of False Data Injection Attacks (FDIA) and Distributed Denial of Service attacks (DDoS). The proposed security model is based on a Markov stochastic process, which is used to observe the behavior of each network device, and employ a range-based behavior sifting policy. Simulation results demonstrate the effectiveness of the proposed architecture and model in detecting and predicting FDIA and DDoS attacks in the context of 5G enabled IoT.
\end{abstract}

\IEEEpeerreviewmaketitle

\section{Introduction}
\label{sec: Introduction}
The rise of 5G networks with the Internet of Things (IoT) systems is set to improve reliable communications and stable connections. 5G new radio access technology has low latency, high availability, and exceptional speed; all needed for several IoT systems \cite{1, o1}. However, beyond providing network performance, 5G enabled IoT systems should also preserve security and improve the reliability of services. A recent report commissioned by the European Union \cite{2} suggests security risks are “likely to appear or become more prominent in 5G networks” because of the extended use of software to run 5G networks. A successfully launched attack in 5G networks may lead to undesired operations and important consequences. This has been understood by hackers who are using new tactics to monetize their attacks by controlling sensitive data, asking for ransom, or making the network unavailable. Yet, not only external hackers could threaten security in 5G networks; insider actors can be the biggest scourges for the system. Network insider actors designate internal workers within the network or network actors. Network insiders with access to sensitive data or resources could cause information breaches or inject false data, for example.

According to 5G PPP, 5G enabled IoT applications are expected to suffer from several security issues due to the complexity and expansion of the attack surface \cite{3, 4}. The increasing scale of devices connectivity and interconnectivity along with network slicing will enable a new range of IoT applications. Yet, one of the weak spots in security will be the devices themselves; they can be remotely controlled to form what is known as a botnet to perform serious security attacks. Most of the existing IoT devices were not developed with security as a priority \cite{icc10}. The volume of data generated by IoT devices is also expected to grow exponentially; traditional intrusion detection methods could be less reliable and further challenging. Therefore, new security attacks detection methods should be further explored in 5G networks. Furthermore, the prediction of network attacks in advance may be a better alternative. Because, once we recognize the possibility of attacks, and verify the insufficient protection, faster mitigation and recovery processes can be deployed.

To protect 5G networks certain security attacks must be addressed; most notably, DDoS attacks. A DDoS attack in the current cellular network (e.g., 4G) can only compromise one service. However, in 5G networks, if a malicious hacker takes control of a slice and launches a DDoS attack, this could compromise services belonging to the same virtual network. In addition, in 5G networks, a DDoS attack could intensify; other installments can also be compromised by this attack if a tunneling protocol is shared between various 5G slices.

IoT systems are enduring an exponential increase of attack surface, because of the large number of possibly compromised IoT devices \cite{icc2, icc1}. Knowing that data transferred by these devices is often critical in decision-making, yet, if it is falsified and quickly transferred using 5G, it could lead to massive consequences for users. Typical 5G enabled IoT systems need some standard security requirements, e.g., 1) resilience to attacks, 2) access control, and 3) data protection. First, resilience to attacks means that the IoT system should not possess a single point of failure, but should adjust itself to network and device failures. Second, data provided by IoT devices requires implementing access control mechanisms to preserve security. For example, before authorizing access to data, user authentication and authorization must be verified. Third, data protection implies the need for resilience to security attacks (e.g., FDIA) and guarantee data authenticity, confidentiality, integrity, and availability.
In this paper, motivated by the aforementioned security issues, we propose 1) a hierarchical architecture for securing 5G enabled IoT networks that uses three layers: the access layer, the multi-access edge computing (MEC) layer, and the cloud layer; 2) a prediction and detection model for FDIA and DDoS attacks based on a Markov stochastic process. The model provides the ability to predict security attacks in situations where the outcome is random, depending only on the present state. Finally, to accurately illustrate the effectiveness of the proposed model, we test it on a healthcare IoT use case.

\section{SECURITY IN 5G ENABLED IOT NETWORKS: RELATED WORK}
\label{sec: Related works}
5G networks will enable IoT devices to communicate and share data faster than ever. However, this development is likely to increase the systems’ vulnerability to security threats, including those from malicious nodes \cite{5}. To overcome some of the security issues, researchers have introduced new solutions that are suitable for 5G networks. For example, to prevent unauthorized action within the network, Xue et al. \cite{6} proposed an efficient access control framework that addresses the problem of a single-point performance bottleneck. The framework uses multiple attribute authorities to partition a load of user legitimacy verification, where each authority manages individually all attributes. Yet, this multi-authority access control framework relies on a central authority responsible for generating secret keys for all users. To ensure key collision resistance between the parallel working authorities, the authors proposed to use timestamp numbers, defined separately in a time interval. Even though this framework is robust for many security issues, it does not ensure resilience to internal system security attacks like FDIA. To address the issue of secure access to slices in 5G networks, Sathi et al. \cite{7} proposed a 5G protocol for securing network slices. The proposed protocol is based on proxy re-encryption, thus ensuring slice isolation and service anonymity.
To mitigate attacks, such as DDoS, which has a huge impact on the network, Sattar et al. \cite{8} proposed a proactive isolation method to mitigate these attacks in 5G core network slices. This isolation method is suitable for inter-slices and intra-slices; however, the scalability requirements underlying IoT applications have not been evaluated in this work. To provide service security in 5G vehicular networks Eiza et al. \cite{9} presented a new system model for vehicular networks, addressing privacy and security issues for real-time video reporting. The proposed cryptographic security model aims to preserve privacy for device-to-device communications while limiting overheads. The model is, however, specific to a vehicular context rather than being a general model for IoT applications in a 5G context.
For various IoT systems, data is collected from heterogeneous devices that support different applications and technologies. This makes the security even more difficult. Recently, several intrusion detection systems have been proposed to secure 5G enabled IoT applications \cite{10, 11}. Ni et al. \cite{12} proposed an efficient secure network-sliced and service-oriented authentication framework for preserving privacy in network slicing and fog computing in 5G networks. The framework enables both 5G stakeholders and IoT systems to support anonymous service-oriented authentication. The authors proposed an access control mechanism based on a key agreement for IoT operators in 5G networks. However, the computational burden and delay brought by the framework may not be appropriate for some IoT applications. Furthermore, to detect intrusion in a network embedded with heterogeneous devices, Loukas et al. \cite{13} proposed an efficient dynamic intrusion system for IoT based on RNN and LSTM to identify DDoS, injection attacks, and other malware. To secure the network and detect intrusions, Fan et al. \cite{14} proposed a hidden Markov model. The issue of this work is that it is not suitable for 5G enabled IoT applications that need real-time responses (e.g., smart hospital).
We can notice that some presented works are not well suited for 5G enabled IoT latency-sensitive applications. Furthermore, some approaches (e.g., \cite{8} and \cite{9}) cannot ensure the scalability needed by IoT systems. Moreover, in 5G enabled IoT systems, data can be manipulated by different operators and actors. Additionally, apart from security attack mitigation, it is essential to focus on attack prediction and detection. In this paper, we propose the adoption of a proactive approach for securing sensitive data by preventing upcoming threats and responding to these threats before they can cause any harm to the systems.

\section{SECURE 5G ENABLED IOT ARCHITECTURE AGAINST FDIA AND DDOS ATTACKS}
\label{sec: system}
In this section, we describe the proposed hierarchical architecture for securing 5G enabled IoT networks. Then, we present the security models implemented in the architecture for the prediction and detection of FDIA and DDoS attacks in 5G enabled IoT networks.
\subsection{A Secure 5G Enabled IoT Architecture: An Overview}
Fig. 1 represents the hierarchical 5G enabled IoT security architecture based on distributed multi-access edge computing (MEC). This architecture uses three layers: the access layer, the MEC layer, and the cloud layer.

\begin{figure*}[t]
	\centering
	\includegraphics[scale=0.48]{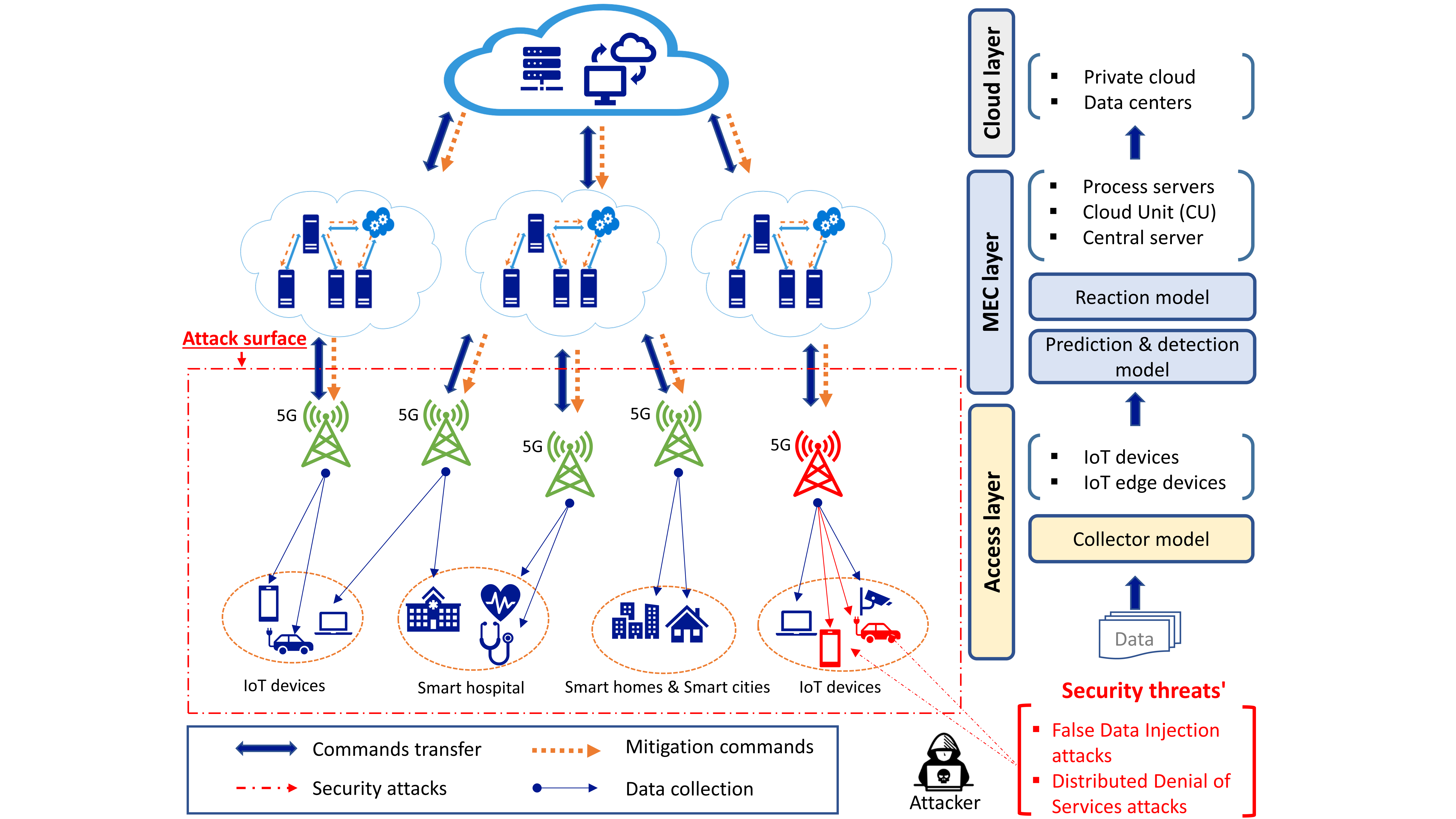}
	\caption{Architecture for 5G-enabled IoT applications: access layer to collect data from the physical environment, MEC layer to predict, detect and mitigate security attacks, and cloud layer to store the data.}
	\label{fig:archi}
\end{figure*}
The access layer contains physical devices that collect and transfer data to the MEC layer. Physical devices could be either servers, access points, or wireless devices. In this layer, all received data are submitted in real time to gateways using the 5G network. The 5G network can provide a high-speed data transfer and a short response time; both necessary for time sensitive and critical IoT applications. The gateways are responsible for managing device connections and forwarding control messages to the corresponding devices. To ensure system scalability regarding the number of connected devices, new gateways can be dynamically activated and managed independently.

The MEC layer processes and analyzes the collected data; it offloads all computational tasks from devices to edge servers to address device limitations like limited computing power and high latency. MEC provides a new ecosystem where communications are rapidly performed between networks using MEC hosts. For example, MEC hosts are usually placed one block or two away from devices, hence communication latency can be low enough to support real-time applications. In this layer, 5G gateways are deployed to aggregate and process the collected data from the access layer and provide supplementary edge services. Although MEC can process a high amount of traffic, it suffers from security challenges. Furthermore, sharing data among multiple devices raises the problem of data leakage and loss. For these reasons, we propose implementing a prediction and detection model (see next section) to mitigate FDIA and DDoS attacks, where a control server belonging to the MEC layer orchestrates communication, sends control messages, and updates security parameters.

The cloud layer manages massive data while the MEC layer manages real-time data generated by devices. In current clouds, a DDoS attack against a resource (e.g., storage, servers, and services) only concerns that resource, and there is a low chance that the attack affects other resources. In 5G networks, DDoS attacks may affect several resources at the same time because they could be tenants of the same shared virtualized infrastructure. A shared cloud among multiple infrastructures requires strict isolation to avoid security leakage and privacy breaches. To mitigate these issues, we propose using a virtual private cloud to limit the risk of data being compromised or altered. The requirements of European security standards prohibit user data disclosure. For example, in the context of healthcare applications, the right to the security of patient medical information used by healthcare professionals must be respected and data confidentiality must be guaranteed. A fundamental security element in the cloud is limiting access to resources and data. We propose to use a cloud access control system to provide users’ security.

To efficiently secure the sensitive data managed by the proposed architecture against FDIA and DDoS, we propose three models: the collector model, the prediction and detection model, and the reaction model. Fig. 2 represents the three models with their components.
\begin{figure*}[t]
	\centering
	\includegraphics[scale=0.48]{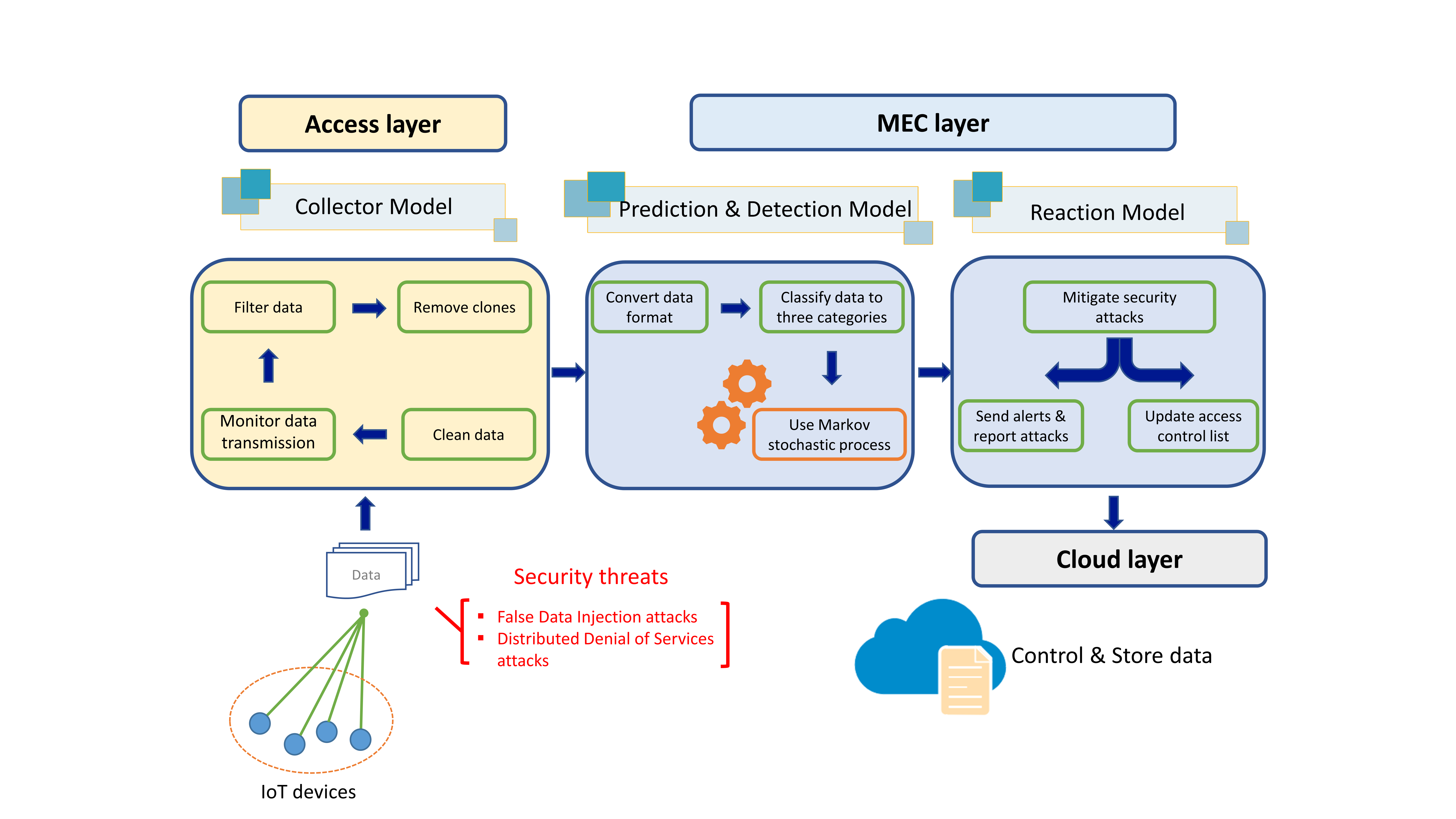}
	\caption{Three models for securing 5G enabled-IoT applications against FDIA and DDoS attacks: collector model, prediction and detection model, and reaction model.}
	\label{fig:archi}
\end{figure*}
\subsection{Prediction and Mitigation of FDIA and DDoS Attacks in 5G Networks}
In the following we will detail the three models used in the proposed architecture to secure 5G enabled IoT systems against FDIA and DDoS attacks:
\begin{enumerate}
    \item \textbf{Collector Model: }
The gateways of the access layer ensure real-time monitoring of devices. They are responsible for gathering all data sent by devices of the access layer and observing whether they transmit or not the sensed data to their destinations. This is realized through a non-selective listening mode, where each gateway treats received data in its communication range. In this module, we implement a cleansing process where we correct and remove inaccurate information. In a 5G context, the data cleansing process should be performed at each gateway to ensure data quality. The ability to collect data while preserving its quality is likely to result in more accurate attack detection. Occasionally, in an IoT scenario data records are duplicated for different reasons. Yet, to avoid system saturation, we propose removing data clones and only keep representative samples; this reduces unnecessary calculations for the prediction and detection of attacks.
\item \textbf{Prediction and Detection Model: }
Data transferred by IoT devices are pre-processed; this step consists in converting the format of records and their classification. Data is often transferred from different devices with heterogeneous configurations; hence, it is vital to convert these data into an understandable and unified format. To support security operations, we classify activities related to data into three categories: reading, updating, and deleting activities. This classification allows an efficient response to security threats alongside data protection. To predict and detect security attacks, we propose a mathematical model based on a Markov stochastic process, where we study the logged activity history of a device. The history log contains information about device states, users’ logins, and system operations, and helps to keep track of system activities based on information and states. Investigating the log history, we can classify security risk over three states: authentic, suspicious, and malicious. An authentic state refers to normal log behavior or a low-risk attack. Suspicious states are detected when the number of logs, i.e., performed actions in the system is higher than a threshold but acceptable by the system. Malicious states happen when the number of logs is higher than a maximal threshold, referring to a high-security risk and possible attacks.
\item \textbf{Reaction Model: }
Building a secure 5G enabled IoT system requires an interconnection between all security mechanisms. For example, devices authentication, access control, and data encryption should be considered together. To mitigate security attacks detected by the prediction and detection model, we implement security mechanisms (e.g., security monitoring and limited access). Fig. 3 presents a  flowchart that details our reaction model that is designed to set security parameters for the detection and mitigation of future attacks, and harden system resilience to attacks. First, we determine the activity’s category. A deleting activity is automatically blocked and the device that tries to perform it is prohibited from the access privileges. Limiting such activity is important because sensitive data must not be lost; old records should be kept for future data analysis. This type of activity is recovered from in a short time. For the two remaining activities (i.e., reading activity and updating activity), we verify the corresponding state. If it is a suspicious state, data are encrypted to protect their integrity and then sent to the cloud layer to be stored. Otherwise, data are sent to an observation state where we evaluate the observation time passed during this step. If this time is higher than the holding time of the current state, we report a malicious behavior that can lead to an attack. The holding time is the amount of time passed before making a state transition, depending on the activity type. The reaction to security threats is related to the state of the system and the security risk (suspicious or malicious). If the behavior is reported as harmful, we block access and monitor the performance of the attack mitigation mechanisms. Otherwise, if it is a harmless behavior, we send control messages to verify the data and update the security settings.
\end{enumerate}
\begin{figure*}[t]
	\centering
	\includegraphics[scale=0.48]{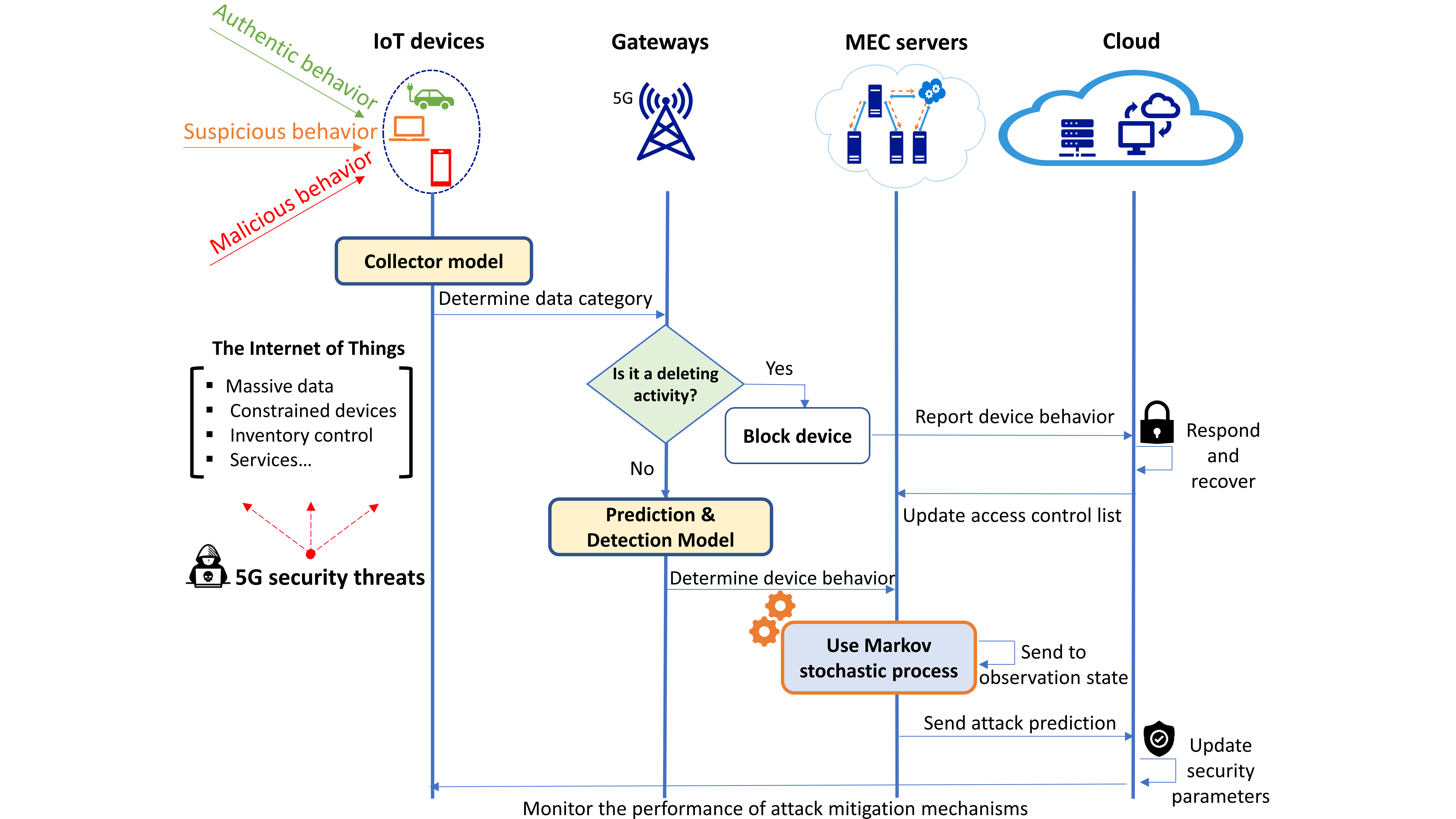}
	\caption{Intrusion detection and reaction strategy deployment.}
	\label{fig:archi}
\end{figure*}

\section{MODEL TO PREDICT AND DETECT FDIA AND DDOS ATTACKS IN 5G ENABLED IOT}
\label{sec: system}
We present a robust stochastic model for FDIA and DDoS attacks prediction and detection based on device behaviors. The behavioral analysis is achieved using a Markov stochastic process for the prediction and detection of attacks.
\begin{figure}[t]
	\centering
	\includegraphics[width=8.2cm, height=7cm]{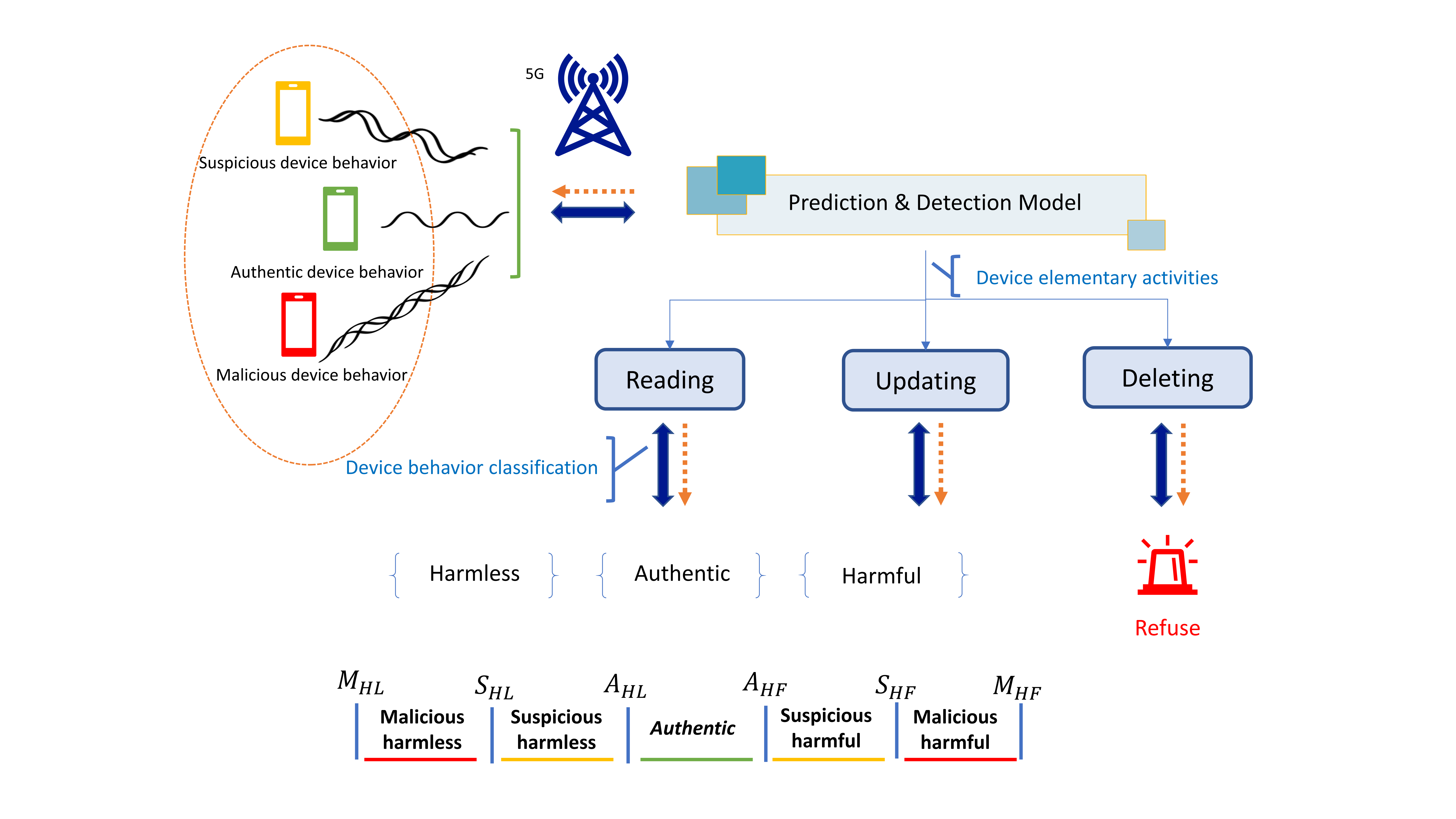}
	\caption{Device state ranges.}
	\label{fig:archi}
\end{figure}
The stochastic modeling aims to introduce various security ranges with different thresholds to identify the behavior of IoT devices in the system based on their log file. The latter records all events that happened between IoT devices and the system. We consider that each device can perform three basic elementary activities which are: reading, updating, and deleting. A device can generate an attack by performing maliciously one of these elementary activities. For instance, a DDoS attack occurs when a device floods the system with too many reading activities. We classify security attacks into two categories: harmless attacks and harmful attacks. A harmless attack can happen accidentally by a device due to mishandling. However, a harmful attack is directed by a malicious device. For each IoT device in the network; the input data into our security model is the number of activities in the log.
We propose a stochastic model to identify the state of a device based on its previous evolution. We use a range-based measurement sifting policy (See Fig. 4) to represent the space of possible states of each device. We aim to study the behavior of each device, then classify it into five classes as shown in Fig. 4 authentic, suspicious harmless, suspicious harmful, malicious harmless, malicious harmful. The main idea of this stochastic model is to introduce five classes with six thresholds values $A_{HL}, A_{HF}, S_{HL}, M_{HL}, S_{HF}$, and $M_{HF}$, corresponding to, respectively, authentic harmless, authentic harmful, suspicious harmless, malicious harmless, suspicious harmful, malicious harmful; and then, we can identify the class of each device. In our work, the thresholds are fixed values calculated using the median of the number of activities in the historical log profile. For example, a device located in the suspicious harmful state at the start of a time interval can move to a malicious harmful state if the number of reported activities are higher than a threshold (i.e., $S_{HF}$).
Due to the space limitation of the paper, we will illustrate only the case of reading activity. For every reading activity, seven states are identified according to the log activity $A_{id}$ corresponding to a time frame, where id is the identifier of a device. The states are listed as follows:

\begin{figure*}[t]
	\centering
	\includegraphics[scale=0.5]{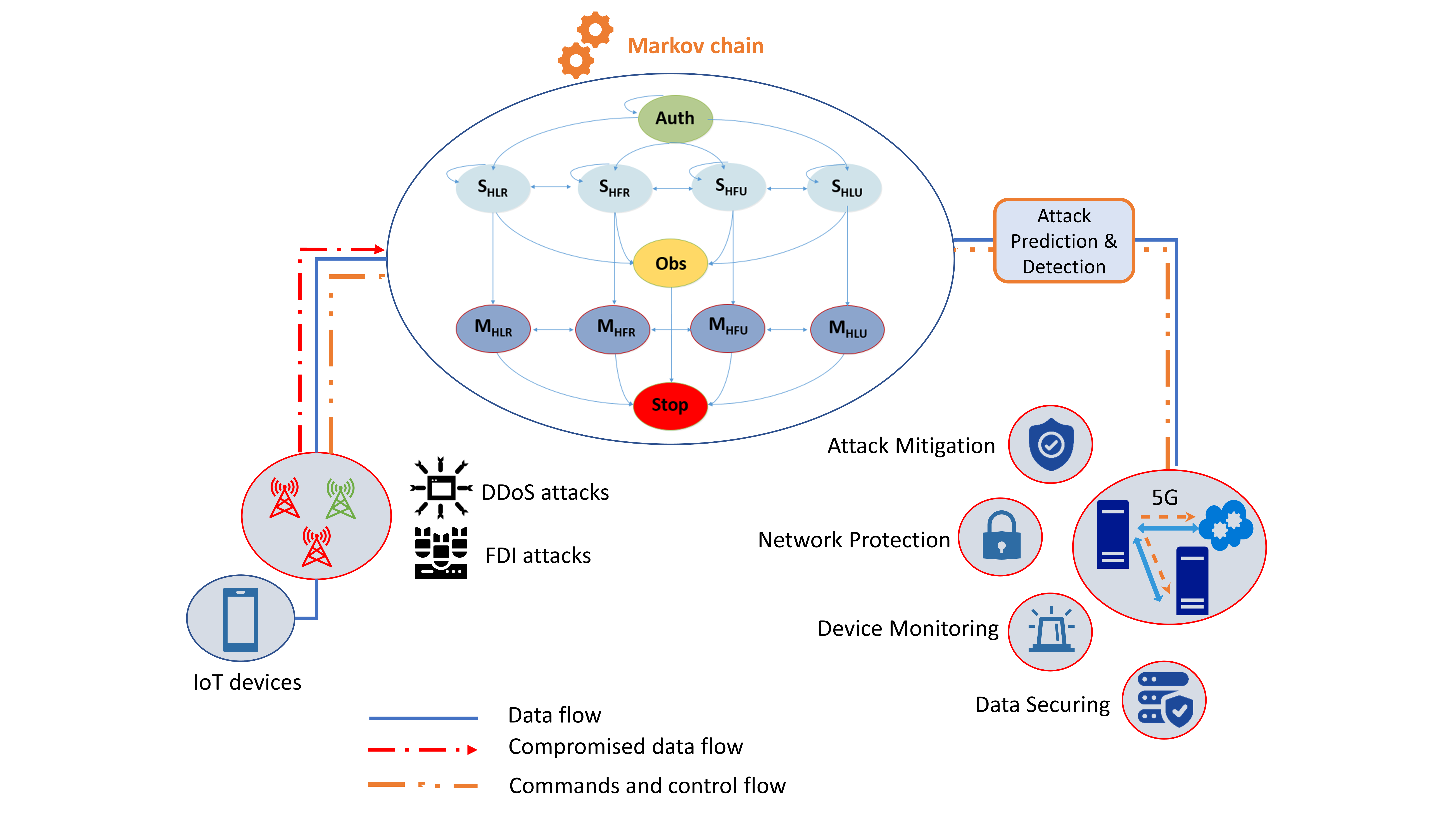}
	\caption{Markov chain process to secure 5G enabled-IoT applications.}
	\label{fig:archi}
\end{figure*}

\begin{itemize}

\item 	Authentic state (i.e., Auth): the device’s number of log activities is between $A_{HLR}$ and $A_{HFR}$.
\item 	Suspicious harmless reading state (i.e., $S_{HLR}$): the device’s number of log activities is between $S_{HLR}$ and $A_{HLR}$.
\item 	Suspicious harmful reading state (i.e., $S_{HFR}$): the device’s number of log activities is between $A_{HFR}$ and $S_{HFR}$.
\item 	Malicious harmless reading state (i.e., $M_{HLR}$): the device’s number of log activities is inferior to $M_{HLR}$.
\item 	Malicious harmful reading state (i.e., $M_{HFR}$): the device’s number of log activities is superior to $M_{HFR}$.
\item 	Stop state (i.e., Stop): a device can be placed in this state when the log activity is within the malicious ranges $M_{HLR}$ or $M_{HFR}$.
\item 	Observation state (i.e., Obs): a device can be placed in this state if it stays during an observation time in one of the two states ($S_{HLR}$ or $S_{HFR}$).
\end{itemize}
Each device can move from one state to another, this is referred to as transition. As shown in Fig. 5, the state space is given by $S_{HLR}$, $S_{HLR}$, $M_{HLR}$, $M_{HFR}$, $S_{HLU}$, $S_{HLU}$, $M_{HLU}$, $M_{HFU}$, Auth, Obs, and Stop. Also, we observe that the Markov chain is irreducible and aperiodic with Semi-process. The evolution of the network is represented by a Semi-Markov process, the state of the network is the union of all devices’ states. Furthermore, the state of a device is memoryless (i.e., the present, the past, and the future are independent). We use a stochastic state transition matrix to represent the evolution of behavior measurement of a device over time. The matrix contains transition probabilities between two states (i.e., classes according to a range-based activity sifting policy), where $P_{i\rightarrow j}$ is the transition probability of a device from state i to state j. The transition probability is the fraction of the number of transitions of a device k from current state i to another state j over the expected number of visits to state i. Each device transition probability denotes the conditional probability that a device transits from one state to another. The Markov process is a random process, i.e., a sequence of random states with a transition probability. For instance, a device can transit from an authentic state (Auth), to a suspicious harmless state ($S_{HL}$), and then transit to an observation state (Obs). However, we can notice in Fig. 5 that a device cannot transit from state Stop to any other states. This state refers to an absorbing state; once a device transits to it, the system is in alert mode. This means that the Markov process determines that an attack has occurred, and an intrusion alert should be reported. All other remaining states are transitioning states.

\section{Evaluation}
\label{sec: evaluation}
\begin{figure*}[t]
	\centering
	\includegraphics[scale=0.5]{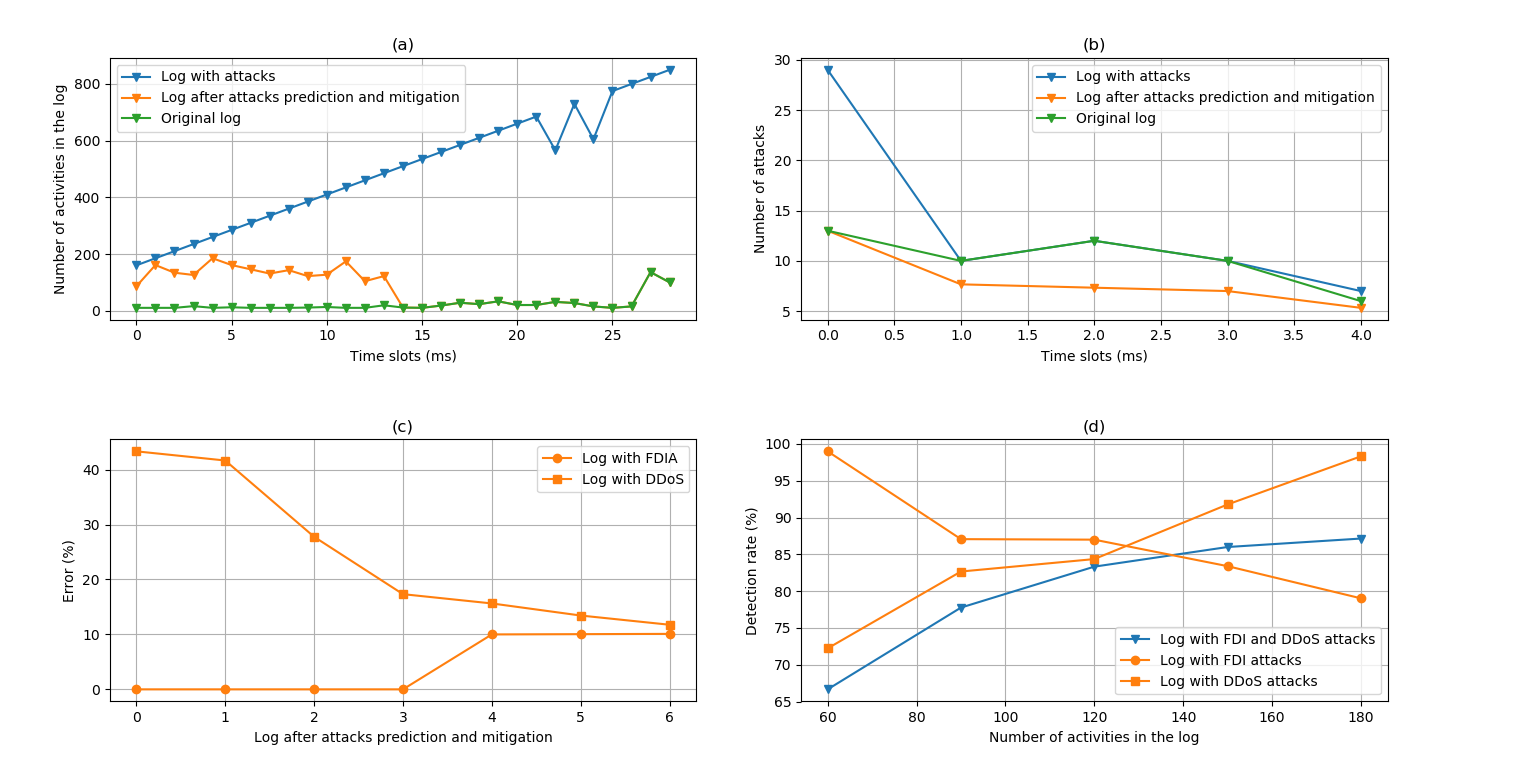}
	\caption{Performance evaluation of a secure 5G enabled-IoT application using Markov stochastic process: a) performance comparison of the number of activities in the log; b) performance comparison of the number of mitigated attacks in the three topologies; c) detection error of FDIA and DDoS attacks in the log after attack prediction and mitigation; d) detection rate of security attacks.}
	\label{fig:archi}
\end{figure*}
In this section, we first describe the evaluation settings, then we discuss the performance evaluation results. To verify the effectiveness of the proposed 5G enabled IoT security solution to predict and mitigate FDIA and DDoS attacks, we performed extensive experiments. The experiments were meant to evaluate the detection rate and the number of attacks mitigated by the proposed security model \cite{icc4, icc3}. We simulated the proposed solution on Intel machine Core™ i7-8550U. This experiment was conducted using real log activity of a mobile health application available online \cite{15}, recording activities performed by the phone sensors. Fig. 6 gives the details of the performance evaluation.
In Fig. 6a, we compare the log activity of three topologies (i.e., log with attacks, log after attacks prediction and mitigation, and the original log without any attacks). In the experiment, we generate the log under attack by randomly generating a number of FDIA and DDoS attacks during a certain time slot. The log after attack prediction and mitigation is the output of the proposed security models within the log with attacks. First, we inject the log under attacks in the collector model to process and clean the data. Then, the prediction and detection model detects malicious behavior and predicts attacks. Finally, the reaction model performs active defense to block anomalous behaviors in the network and mitigate the detected attacks. The original log is the log extract from the healthcare application. In Fig. 6a, we can observe that the proposed security solution mitigates all attacks after some time. Furthermore, after a period we observe that the log after attacks prediction and mitigation and the original log are overlapped; this explains the efficiency of the proposed models to mitigates attacks.

In Fig. 6b, we compare the number of attenuated attacks as a function of time. To assess the reaction model and its ability to remember old attacks, we compare the secure log and original log with the log under attack. The log with attacks is made by reducing the number of attacks over time. From this figure, we observe that the proposed solution mitigates the maximum number of attacks. Besides, even if the original log is considered to illustrate a secure system, we notice that certain attacks were not detected while the proposed security solution attenuated them.

In Fig. 6c, we evaluate the intrusion error rate of the security model against FDIA and DDoS attacks. We injected a fixed number of attacks to the original log and calculated the error rate. We observe that our solution is effective when the number of activities is small in time; the error rate of prediction FDIA increases with the number of logs. Yet, this error increases slowly. With further log activities, we observe that the rates of error detection of FDIA and DDoS attacks are reduced and almost equal.

In Fig. 6d, we randomly generate fifty attacks and evaluate the detection rate of our security solution. We observe that while increasing the number of log activity, the detection rate of the prediction and detection model increases. This figure displays the scalability of our solution and its capacity to predict and detect security attacks when having many logs.

\section{DISCUSSIONS AND FUTURE DIRECTIONS}
\label{sec: evaluation}
This section presents a discussion of the proposed security model and the possible future direction.
To efficiently achieve security and resilience to the ever-evolving threat landscape in 5G enabled IoT networks, we proposed a security model to predict and detect malicious device behavior that can cause FDIA and DDoS attacks. To predict the occurrence of an attack in the network, the security model uses a Markov process. This prediction can be made regarding the occurrence of an attack in the network, based only on its current state without knowing the log history. Indeed, the proposed model predicts attacks by a device by only knowing the current state of the device; it does not need the full-size log history file to predict an attack. This allows both a swift detection/mitigation process, and overcoming the problem of the exponential growth of log size. The Markov process is an analytical method, which means that the reliability parameter of the prediction is calculated using a probability formula. This has a considerable advantage of uncovering future attacks and bringing faster and accurate mitigation to the network. Indeed, the early prediction of potential attacks in the network can lead to an efficient reaction to them. Furthermore, the Markov process is lightweight and does not require complex computing; it can be deployed at the edge of the network.

To detect attacks in the network, we propose tracking device activities in the network. Therefore, we proposed using a range-based activity sifting policy based on different thresholds to represent the space of possible device states. The state of a device corresponds to its current activity in the network. We consider that each device can transit from one state to another, and to identify each state we used fixed threshold values determined from the historical log profile of the device. For example, we can detect a false data injection attack when the number of activities of a device during a period is higher than the permitted thresholds. In this work, the detection of attacks is conducted with fixed threshold values. However, we consider that the threshold values determination for each device should be integrated dynamically into the Markov process. In our future work, we intend to evaluate the use of a reinforcement learning strategy to determine these threshold values for each device in the network.

\section{Conclusion}
\label{sec: Conclusion}
In this article, we presented a hierarchical architecture for securing 5G enabled IoT networks, and a stochastic Markov model for securing against FDIA and DDoS attacks. The architecture includes three tiers (i.e., the access layer, the MEC layer, and the cloud layer) and implements three security models (i.e., collector models, prediction and detection model, and reaction model). We have detailed the stochastic Markov detection and prediction model to mitigate FDIA and DDoS attacks by examining network devices behavior. Finally, we have evaluated the performance of the proposed security solution using a healthcare application as a use case. The extensive simulation results showed a low error rate, a high detection rate, and a decrease in the number of attacks in a short period.
\section*{acknowledgement}
\label{sec:acknowledgement}
The authors would like to thank the Natural Sciences and Engineering Research Council of Canada, as well as FEDER and GrandEst Region in France, for the financial support of this research.

\label{Related work}
\bibliographystyle{IEEEtran}
\bibliography{./references}


\end{document}